\documentclass[sigconf]{acmart} 
\AtBeginDocument{%
  }

\setcopyright{acmlicensed}
\copyrightyear{2026}
\acmYear{2026}

\acmConference[DIS Companion '26]{Designing Interactive Systems Conference}{June 13--17, 2026}{Singapore, Singapore}

\usepackage{subfiles}
\usepackage{multirow}
\usepackage{graphicx}
\usepackage{subcaption}
\usepackage{listings}
\usepackage{pgfplots}
\usepackage{pgf-pie}
\usepackage{soul}
\pgfplotsset{compat=1.17}
\usepackage{array}
\usepackage{float}
\usepackage{csquotes}
\usepackage{xcolor}
\usepackage[most]{tcolorbox}
\usepackage{CJKutf8}
\usepackage{booktabs}
\usepackage{tabularx}
\usepackage{longtable}
\usepackage{pdflscape}
\usepackage{pdfpages}
\usepackage{makecell}
\usepackage{tabulary}
\usepackage[table]{xcolor}
\usepackage{arydshln}
\usepackage{pifont}
\usepackage{circledsteps}

\newcolumntype{P}[1]{>{\raggedright\arraybackslash}p{#1}}
\newcolumntype{L}{>{\raggedright\arraybackslash}T}

\newif\ifredact
\redacttrue   

\newif\ifcomment
\commenttrue   
\ifcomment
  \newcommand{\missing}[2][]{\textcolor{red}{[\textbf{MISSING\ifx#1\empty\else~–~#1\fi}] ~#2}}
  \newcommand{\kel}[1]{~\sethlcolor{pink}\hl{[Kellie: #1]}}
  \newcommand{\ken}[1]{~\sethlcolor{yellow}\hl{[Kenny: #1]}}
  \newcommand{\pinsym}[1]{~\sethlcolor{cyan}\hl{[Pin Sym: #1]}}
  \newcommand{\darryl}[1]{~\sethlcolor{lightgray}\hl{[Darryl: #1]}}
  \newcommand{\johnhenry}[1]{~\sethlcolor{green}\hl{[John-Henry: #1}}
\else
  \newcommand{\missing}[2]{}
  \newcommand{\kel}[1]{}
  \newcommand{\ken}[1]{}
  \newcommand{\pinsym}[1]{}
  \newcommand{\darryl}[1]{}
  \newcommand{\johnhenry}[1]{}
\fi

\newcommand{\acpagent}[1]{\textsc{ACPAgent}}
\newcommand{\baritone}[1]{\textbf{\textsf{P1}}}
\newcommand{\redpasta}[1]{\textbf{\textsf{P2}}}
\newcommand{\trombone}[1]{\textbf{\textsf{P3}}}
\newcommand{\cowalive}[1]{\textbf{\textsf{P4}}}
\newcommand{\grootfox}[1]{\textbf{\textsf{P5}}}
\newcommand{\seaonion}[1]{\textbf{\textsf{P6}}}
\newcommand{\sticklog}[1]{\textbf{\textsf{P7}}}
\newcommand{\donutweb}[1]{\textbf{\textsf{P8}}}
\newcommand{\fogalive}[1]{\textbf{\textsf{P9}}}
\newcommand{\starship}[1]{\textbf{\textsf{P10}}}
\newcommand{\toystory}[1]{\textbf{\textsf{P11}}}
\newcommand{\goldfish}[1]{\textbf{\textsf{P12}}}
\newcommand{\pumpkins}[1]{\textbf{\textsf{P13}}}
\newcommand{\sunsalad}[1]{\textbf{\textsf{P14}}}
\newcommand{\webzebra}[1]{\textbf{\textsf{P15}}}

\newcommand{\dquote}[1]{\enquote{#1}}

\begin{document}
\renewcommand\footnotetextcopyrightpermission[1]{} 
\settopmatter{printacmref=false} 

\title[Rethinking Provenance, Temporality and Legitimacy in Post-Mortem Agents]{Acts of Configuration: Rethinking Provenance, Temporality and Legitimacy in Post-Mortem Agents}


\author{Kellie Yu Hui Sim}
\email{kellie_sim@mymail.sutd.edu.sg}
\orcid{0009-0005-6451-7089}
\affiliation{
  \institution{Singapore University of Technology and Design}
  \city{Singapore}
  \country{Singapore}
}
\authornote{Corresponding author. This is a preprint of the paper accepted at DIS 2026. The final version will be available in the ACM Digital Library.}

\author{Pin Sym Foong}
\email{pinsym@nus.edu.sg}
\orcid{0000-0002-4437-8326}
\affiliation{
  \institution{Telehealth Core \\ National University of Singapore}
  \city{Singapore}
  \country{Singapore}
}

\author{Darryl Lim}
\email{darryl_lim@mymail.sutd.edu.sg}
\orcid{0009-0007-8173-2199}
\affiliation{
  \institution{Singapore University of Technology and Design}
  \country{Singapore}
}

\author{John-Henry Lim}
\email{lim_johnhenry@mymail.sutd.edu.sg}
\orcid{0009-0000-2168-9722}
\affiliation{
  \institution{Singapore University of Technology and Design}
  \country{Singapore}
}

\author{Kenny Tsu Wei Choo}
\email{kenny_choo@sutd.edu.sg}
\orcid{0000-0003-3845-9143}
\affiliation{
  \institution{Singapore University of Technology and Design}
  \country{Singapore}
}

\renewcommand{\shortauthors}{Sim et al.}

\begin{abstract}
Work on persona-persistent post-mortem agents typically frames design around a life/death binary.
This framing neglects a consequential yet under-theorised condition: when individuals remain alive but have impaired decisional capacity.
Drawing on a multi-phase workshop in which participants trained and reflected on an AI agent for Advance Care Planning, we examined how people reason about agentic delegation post-capacity loss.
Initially, participants favoured bounded agents grounded in first-party authorship and representational fidelity over autonomous or evolving stand-ins.
However, temporality introduced novel ideas like adjacent use driven by persona persistence over functional expansion: agents should evolve while users retain capacity, remain static once capacity is lost, but somehow inform adjacent post-mortem uses. 
We discuss the implications of these findings and propose that the configuration of agents for post-capacity use reshapes our understanding of provenance, temporality, and legitimacy for post-mortem agents.
\end{abstract}



\maketitle

\section{Introduction}
Research on agents that create persona persistence after death has many names, such as griefbots \cite{HollanekNowaczyk2024Deadbots}, post-mortem agents \cite{morrisGenerativeGhostsAnticipating2025}, deadbots \cite{HollanekNowaczyk2024Deadbots}, and so on.  
However, literature tends to assume a binary framing: pre- or post-mortem. For example, a systematic review of related papers in HCI framed the outcomes as systems for 'dying' and for 'death'  ~\cite{albers_dying_2023, foongDesigningCaregiverfacingValues2024, sim_words_2025}.  
In contrast, there also exists research on systems that are designed for use after an individual has lost cognitive capacity, but is still alive.  
In this liminal stage between life and death, the individual's participation is uncertain. They remain alive but variably present. Therefore, it cannot be categorised as being in the post-mortem phase. 
This body of work, found in health \cite{fooBenefitsRisksLLMs2025} and in financial application areas~\cite{latulipeUnofficialProxiesHow2022}, is motivated by the need to delegate decisional authority in preparation for the loss of decisional capacity associated with age- or illness-linked impairments. 
Such past work targets a liminal space of individuals who are alive but have impaired decision-making as they approach death. 
For this paper, we will label this period as "post-capacity", placing it on a timeline between pre- and post-mortem design. 

To our knowledge, this phase has not been examined as a distinct design space, particularly in relation to both the "alive and well" phase and post-mortem systems. Existing frameworks (e.g., generative ghosts~\cite{morrisGenerativeGhostsAnticipating2025}), primarily assume post-mortem deployment and do not examine how pre-death configurations affect post-mortem representations or provenance. We argue that treating post-capacity as merely an extension of pre-mortem agency risks obscuring the unique tensions that emerge when further adaptation becomes uncertain or impossible.
Studying the post-capacity period is important because decisions about values, delegation, and boundaries made while individuals retain capacity shape how representation unfolds during cognitive decline and after death.
Further, systems developed before or during cognitive impairment may later function as artefacts or as memorabilia for the bereaved, influencing perceptions of authenticity and responsibility.
By conceptualising post-capacity as its own design space, we move beyond a binary life/death framing and foreground how agency and accountability evolve across states of capacity.
Questions of \textit{provenance} (authorship and control) and \textit{temporality} of the arrangements for adaptability and responsibility may become especially fraught when an agent is expected to act on one's behalf without the possibility of correction, revision, or consent. 

Recent work has explored post-capacity AI agent framings using an experience prototype in the Advance Care Planning (ACP) domain to simulate agent training~\cite{sim_words_2025}.
The study concluded that laypersons simultaneously and differently conceive of AI agents for high-subjectivity decision-making as educational tools, as patient advocates speaking on their behalf, or as legal proxies making decisions for them.
Across these, \emph{advocacy} emerged as the most promising framing when the \textit{source persona}~\cite{Lei2025AIAfterlife} is no longer able to speak for themselves. 
Building on the idea of advocacy, this provocation examines unanswered questions about the design of agent advocates, and their boundaries and evolution when correction or consent cannot be assumed. 
We examine the link between \textbf{prior development and initiation to post-mortem parameters of provenance and temporality} through the study of designing for post-capacity use.
In doing so, we provide situated qualitative insights into how participants reasoned about agent-mediated delegation, rather than statistically generalisable claims.

\section{Methodology}
We conducted a 2.5-hour, multi-phase workshop study that received Institutional Review Board approval (SUTD IRB-25-00727). All procedures followed ethical research guidelines.
We recruited 15 participants (7 female; aged 40 and above, $M=53.1$, $SD=7.9$) via social media, targeting individuals with personal/family experience of life-limiting illness; all were compensated $\sim$USD\$23.
Most had prior exposure to our country's ACP system (13/15); 6 had acted as donors/proxies, and 6 could not identify a trusted proxy.
13 reported close family/friends with life-limiting illness, and 11 had used digital or AI tools for health-related decision-making.

\begin{figure*}[htbp!]
    \centering
    \begin{subfigure}[t]{0.45\textwidth}
    \includegraphics[width=\linewidth]{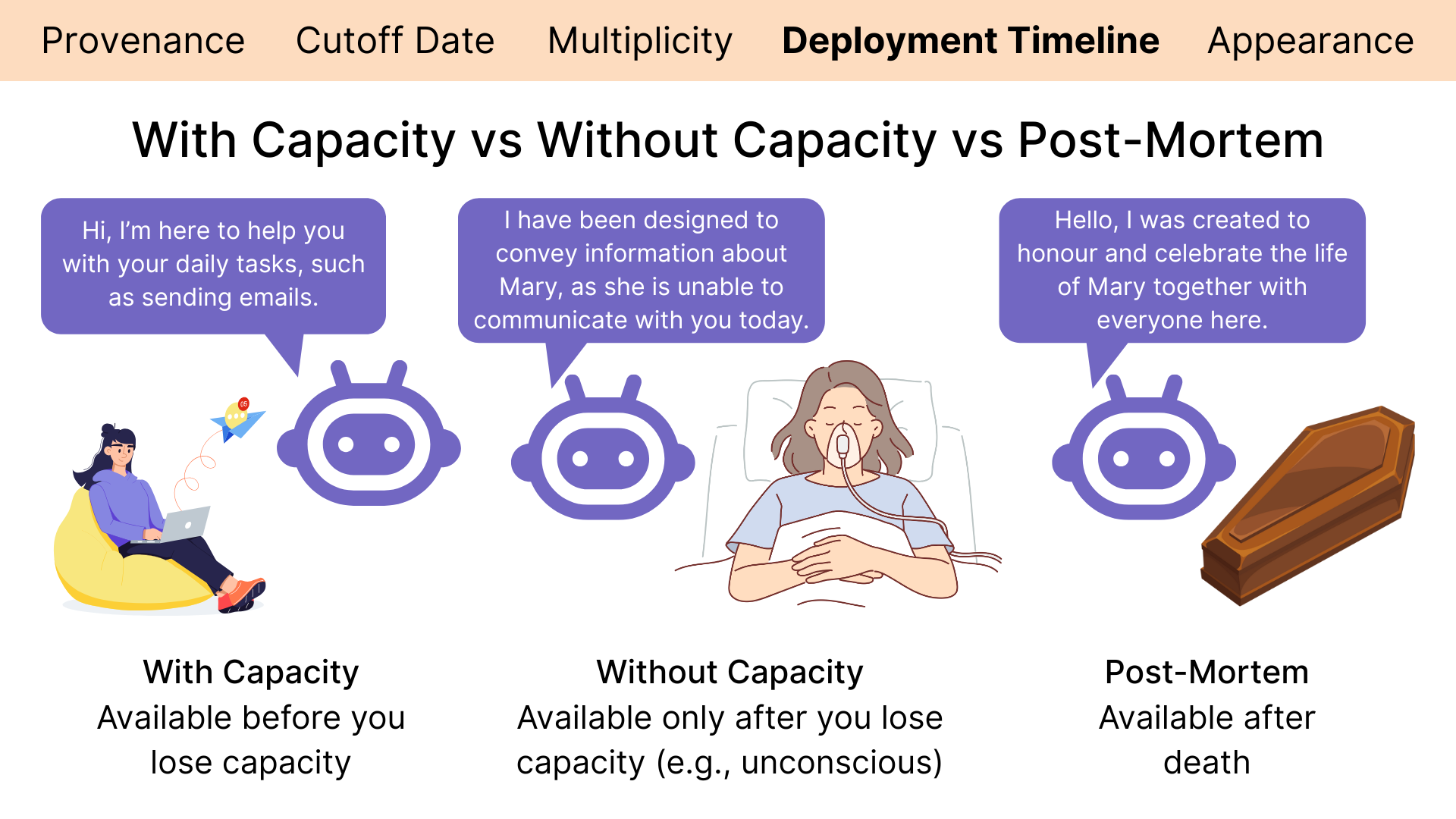}
    \caption{Illustration of capacity, without capacity, and post-mortem.}
    \Description{A three-panel illustration showing different deployment timelines for an AI agent. On the left, a person using a laptop is paired with an agent labelled "With Capacity," indicating availability before the loss of capacity. In the centre, a hospitalised individual is paired with an agent labelled "Without Capacity", indicating availability only after loss of capacity. On the right, a coffin icon is paired with an agent labelled "Post-Mortem", indicating availability after death.}
    \label{fig:deployment}
    \end{subfigure}
    \hspace{0.08\textwidth}
    \begin{subfigure}[t]{0.45\textwidth}
    \includegraphics[width=\linewidth]{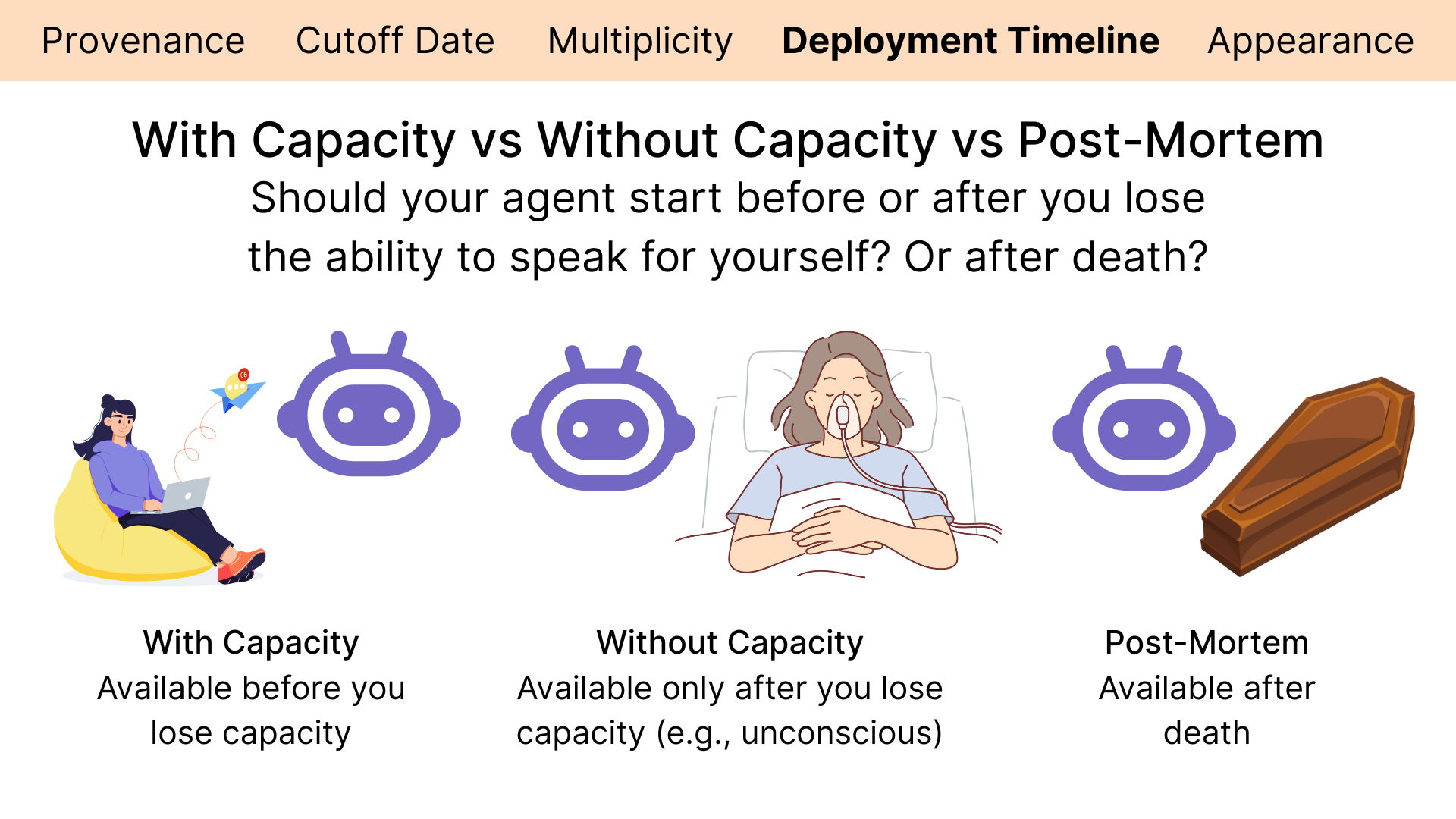}
    \caption{Slide reframed as a prompt.}
    \Description{An illustration similar to the previous figure, again depicting three deployment timelines: with capacity, without capacity, and post-mortem. The header asks whether an agent should start before, after, or after the loss of the ability to speak for oneself, or after death, reinforcing deployment timing as a design decision.}
    \label{fig:deployment-prompt}
    \end{subfigure}
    \caption{Slides used to introduce and prompt discussion of the \textbf{deployment timeline} parameter (fourth parameter).}
    \label{fig:parameter_slides}
\end{figure*}

Earlier phases examined participants' hands-on training of an LLM-enhanced agent for decision support in ACP~\cite{sim_words_2025}.
Results informed design requirements for future care planning agents and identified a potentially viable framing of this agent as an advocate.
This provocation focuses on the results of the subsequent structured co-design discussion, examining how participants reasoned about provenance and temporality when agents might represent their values after loss of decisional capacity in high-risk contexts.
Here, the discussion was guided by slides illustrating parameters inspired by the generative ghosts framework~\cite{morrisGenerativeGhostsAnticipating2025} (see Figure~\ref{fig:parameter_slides}).
Participants responded to each parameter prompt by discussing how they would configure and constrain an agent to act on their behalf, often drawing on their prior interaction with the system~\cite{sim_words_2025}.
These discussions surfaced participants' reasoning about authorship (e.g., who should define decisions), adaptability (e.g., whether agents should evolve), and scope (e.g., when agents should act or stop).
We treated these responses as situated articulations of design expectations, which formed the primary data for analysis and informed our findings.
Sessions were audio-recorded and transcribed using WhisperX~\cite{bain2022whisperx}.

Using framework analysis~\cite{galeUsingFrameworkMethod2013}, three researchers independently coded transcripts, beginning with deductive codes derived from workshop parameters and tracking excerpts for contextual grounding.
Through iterative team discussions and open coding, we collapsed codes across and within parameters to generate higher-level themes.
The analysis adopted a contextualist stance, focusing on semantic meaning while recognising the researchers' interpretive role.

\section{Results}


\subsection{Boundedness of Advocate Agents}
Participants found AI agents most acceptable when framed as bounded advocates, whose legitimacy derived from faithfully conveying previously articulated human intent rather than deciding, optimising, or interpreting preferences.
Agents positioned as surrogate decision-makers that sought to determine the \dquote{right} choice were less acceptable.
Acceptable agency was understood as retrieving, communicating, or reformulating what had already been expressed in response to queries from the agent's users (i.e., decision stakeholders), so that others would \dquote{know what is my wishes} (\seaonion{}).

Most participants favoured first-party authorship, emphasising that ACP decisions should be authored by the individual who provided the source persona data.
\pumpkins{} was adamant on being the key decision driver for his own agent, viewing it as the only sensible thing to do, stating, \dquote{I will be the decision maker. Nobody will leave the decision to the third party}.
Even when participants were open to third-party involvement, such involvement was conditional and supportive rather than substitutive.
Third parties were described as intermediaries who could help reduce emotional burden or facilitate articulation, but not as authors or interpreters of values.
For example, \toystory{} noted that in making a decision for her mum, she would want to do it \dquote{together with her with someone facilitating [...]}, as she did not \dquote{want to just go and make decisions that I feel is right for her [...] So I want someone to facilitate that kind of situation}.

This emphasis on boundedness was closely linked to how participants reasoned, trusting the agent to act, which was grounded in stability rather than in intelligence or correctness.
Participants repeatedly stressed that the agent should not reinterpret values or evolve autonomously in ways that could introduce interpretive drift, where the agent moves beyond explicitly stated preferences and begins to infer or extend them.
As \trombone{} explained, he preferred a static agent (i.e., one that would never change after deployment approved by them) because he was \dquote{not asking [the agent] to make the right decision [...] but to make my decision}. 
In this sense, constraint functioned as a trust mechanism: limiting what the agent could do was seen as essential to entrusting it with accurate representation.

Boundedness of the agent was also evidenced by participants' preference for a single, integrated agent over multiple specialised ones.
Rather than distributing responsibilities across separate agents for ACP, legal matters, or personal affairs, participants described a desire for a \dquote{one-stop} system that consolidates information, reflects consistent values, and provides structured guidance across domains.
This preference for limited functionality was not due to distrust of technical capability, but to reduce emotional and cognitive burden during crisis situations.
In moments of stress or diminished capacity, navigating different platforms, repeating information, or reconciling conflicting outputs was overwhelming, whereas a unified agent could streamline decision-making and minimise frustration.

Consequently, participants treated visual representation as secondary to function, often favouring minimal or clearly artificial (i.e., non-human) designs.
Highly anthropomorphic representations were seen as risking over-identification or misplaced expectations of agency.
Together, perceived trustworthiness in agents seemed to be shaped more by clear limits, representational accuracy, and cognitive simplicity than by evolving, intelligent, or high-fidelity agents.


\subsection{How Temporality Counters Agent Bounding}

Given the emphasis on boundedness, it is unsurprising that our participants initially struggled to envision how post-capacity agents might be used in post-mortem contexts. 
However, upon viewing the timeline slide illustrating with/without decisional capacity and post-mortem (Fig.~\ref{fig:parameter_slides}), participants began to connect elements of post-capacity agents to post-mortem usage.
By introducing temporality, the slide clarified expectations during periods with capacity versus after its loss, and helped participants articulate how agents might transform across phases.

While decisional capacity remained, participants wanted agents to remain malleable by supporting the ongoing authorship and refinement of preferences.
They expected agents to help articulate, revise, and confirm their values over time, emphasising transparency, editability, and understanding.
In this phase, trustworthiness was grounded in the agent's ability to \textit{support provenance behaviour such as human authorship} rather than replace it.

The transition from post-capacity to pre-mortem led to ideas of agent multiplicity and the permissibility of adjacent use. 
When participants were comfortable with the agent being multi-purpose (i.e, \dquote{a one-stop service}), they added the proviso that its use remained within domains they perceived as "closely related" to ACP.
In this case, multiplicity was seen as supporting adjacent tasks, such as matters related to LPAs, wills, or digital legacy planning, rather than as a general-purpose capability (e.g., spanning unrelated domains such as everyday or work-related tasks). 
These extensions were not framed as an \textit{expansion} of authority, but as a way to \textit{maintain continuity} of the source persona across closely related domains and reduce the emotional burden of repeated articulation.
However, what counted as \emph{adjacent} varied substantially across participants and was rarely articulated consistently. 
This reveals an unresolved tension for designers of agentic AI: even when users prefer bounded agents, the practical boundaries of that purpose remain unclear, and it is uncertain how we should define the emerging concept of \textit{task adjacency}.

Furthermore, adjacency of purpose did not extend well from a temporal perspective. 
Participants distinguished future and post-mortem use, arguing that ACP agents should end once care preferences were conveyed (\dquote{it ends the chapter}).
If retained post-mortem, its function ought to shift.
Instead, participants suggested that the agent should now be associated with commemoration or reassurance rather than decision representation.
While \toystory{} cautioned that lifelike representations or voice recordings might intensify grief and should remain optional with appropriate support, \baritone{} saw value in archived messages or simulated reactions that could offer reassurance during bereavement but emphasised separation because \dquote{the post-mortem should be a separate product [...] shouldn't mix with the ACP}.

Across reflections, the source persona's decisional capacity emerged not just as contextual background but as a structuring boundary that determined when agents could author input, re-present information, or withdraw from further action.
Participants' trust in agent advocates depended on systems respecting  temporal boundaries: evolving when users had capacity, remaining static when they did not, and ceasing once advocacy was no longer appropriate.

\section{Discussion, Limitations and Future Work}

\subsection{The Act of Configuration Influencing Post-Mortem Agent Autonomy}
Our findings suggest that participants preferred agents that acted as bounded advocates, faithfully conveying previously articulated intent.
Trust and acceptability emerged not as a function of what the agent could do, but of what it was explicitly constrained to do. Temporality introduced options for expanding the boundaries, but only to adjacent domains in order to maximise the utility of the persona agent. 

Extending prior work on generative ghosts~\cite{morrisGenerativeGhostsAnticipating2025}, our findings suggest that the source persona's involvement in agent configuration does not simply enable post-mortem persistence, but actively constrains it.
While research on AI replicas, proxies, or self-clones often emphasises continuity of identity, realism, or extended personhood~\cite{karpusPersonsTheirDigital2025, huangMirrorCompanionExploring2025}, our participants reasoned about delegation more conservatively.
Instead of offloading responsibility or creating autonomous stand-ins, they emphasised preserving authorship across time, particularly in situations where future correction or intervention by the source persona would no longer be possible.
First-party provenance and static or explicitly bounded behaviour were preferred not due to maximised capability, but due to the reduced risk of interpretive drift, where the system moves beyond stated preferences.
In this sense, trust was grounded less in autonomy or realism than in restraint and representational fidelity.

We conclude that when individuals are directly involved in training and bounding agents, legitimacy becomes anchored not only in data fidelity but in conditions of initiation and authorisation.
This emphasises post-mortem agent autonomy as contingent on earlier acts of configuration instead of as an independent phase of deployment.


\subsection{The Challenge of "Adjacent" Use}
The findings on temporally driven adjacent use offer nuance to the idea of bounded single-purpose agents. 
Recall that adjacency of use was connected to \textit{persona} continuity, but not \textit{temporal} continuity. 
However, what counted as \emph{adjacent} tasks varied substantially across participants and was rarely articulated in consistent terms. Participants viewed multiplicity as support for adjacent tasks, such as matters related to LPAs, wills, or digital legacy planning, rather than as a general-purpose capability (e.g., spanning unrelated domains such as everyday or work-related tasks).
These extensions were not framed as an expansion of authority, but as a way to maintain continuity of values across closely related domains and reduce the emotional burden of repeated articulation.
One possible response to this ambiguity is to distribute adjacent functions across multiple coordinated agents.
However, emerging work on multi-agent systems suggests that agent collectives may exert social or normative influence on users, shaping decisions through perceived group norms or a desire to avoid discord rather than through explicit reasoning~\cite{songMultiAgentsAreSocial2025, songGreaterSumIts2025}.
Furthermore, how such agents would coordinate with one another and how users would appraise and interpret their interactions remains the subject of future work.
From this perspective, introducing multiplicity may reconfigure how authority and control are experienced, underscoring the need for careful scoping and governance when distributing responsibilities across agents.

Together, our findings reveal a design tension: bounding agents raises questions of scope and adjacency, yet extending them risks unendorsed growth, unpredictable futures, and agentic action without the possibility of correction.

\subsection{Recognising the Post-Capacity, Pre-Mortem Design Space}
Our findings foreground post-capacity, pre-mortem agency as a distinct phase in which agency is neither fully active nor fully concluded.
Instead of using this phase as an extension of deliberation when fully in possession of function, or as a precursor to post-mortem representation, participants articulated clear expectations for use, transition and termination in the post-capacity phase.
Agents were expected to evolve while users retained capacity, remain static once capacity was lost, and withdraw once their advocacy role was fulfilled.
Unlike speculative or normative accounts of post-mortem agents~\cite{morrisGenerativeGhostsAnticipating2025, HollanekNowaczyk2024Deadbots}, participants in our study acted as the \textit{source data} and \textit{creator}, and therefore could ground their reasoning in their personal and recent experience of training an agent.
Thus, their desire for an agent bounded in time and function lays a clear emphasis on mistrust of free-acting, evolving self-representations in deadbots and post-mortem agents. 
Whether this preference lies in perfect contravention of the acceptability deadbots remains the subject of future work. 
Certainly, this challenges assumptions that continuity or ongoing learning is inherently desirable in post-mortem systems, particularly when future correction is impossible. Hence, future work might further explore how structured training or preparatory interactions support reasoning about post-capacity delegation, as well as how these patterns extend across larger and more diverse populations, including different cultural contexts and care settings.



\section{Conclusion}
This study examined people's reasoning about AI agents intended to act on their behalf in anticipation of decisional capacity loss.
In contrast to previous work, participants preferred tightly bounded agents, prioritising constraint, representational fidelity, and authorship preservation over autonomy or intelligence.
They expected agents to support articulation while capacity remained, retrieve prior intent without reinterpretation once capacity was lost, and withdraw when advocacy was complete.
They resisted conflating agents with post-mortem or memorial systems, emphasising clear transitions and endpoints, but were open to adjacent uses.
We conclude that the value of agent advocates in high-risk, high-subjectivity contexts lies not in autonomy or persistence, but in constrained representation that respects temporal boundaries and authorship.
By examining participants' experiences of training an agent for post-capacity use, this work grounds scholarship on delegation, temporality, and responsibility in human-AI interaction.
Together, these findings position the post-capacity, pre-mortem phase as a distinct and under-explored design space.
\begin{acks}
We thank Natasha Ureyang and Huynh Vinh Anh for their valuable inputs into the design of the study.
We also thank Chenyu Zhao and Swarangi Subodh Mehta for their support in facilitating the workshop study.
This work was partially supported by the Singapore Ministry of Health’s National Medical Research Council under grant numbers NMRC/CG1/009/2022-NUH and CareEco21-0030.
We are deeply grateful to our participants for sharing their time, experiences, and candid reflections with us.
Finally, we thank our reviewers for their thoughtful feedback and constructive suggestions, which helped strengthen this work.
\end{acks}

\bibliographystyle{ACM-Reference-Format} 
\bibliography{main}



\end{document}